\documentclass[draft]{agujournal2019}
\usepackage{url} 
\usepackage{outlines}
\usepackage[utf8]{inputenc}
\usepackage[inline]{trackchanges} 
\usepackage{soul}
\draftfalse

\journalname{Geophysical Research Letters}

\begin{document}

%
%


\title{Background Pycnocline depth constrains Future Ocean Heat Uptake Efficiency}

%
%




\authors{Emily Newsom\affil{1}, Laure Zanna\affil{1}, Jonathan Gregory\affil{2}\affil{3}}

 \affiliation{1}{Courant Institute of Mathematical Sciences, New York University}
\affiliation{2}{National Centre for Atmospheric Science, University of Reading, Reading, UK}
 \affiliation{3}{Met Office Hadley Centre, Exeter, UK}





\correspondingauthor{=name=}{=email address=}




\begin{keypoints}

\item Pycnocline depth correlates strongly with OHUE in CMIP5/CMIP6 and MITgcm.
\item A regional decomposition of OHUE illustrates that mid-latitude heat uptake and sequestration drives the correlation between OHUE and pycnocline depth. 
\item Inter-model differences in pycnocline depth explain around 70$\%$ of the spread in OHUE across CMIP5 and CMIP6.

\end{keypoints}

%
%

%
%


\begin{abstract}
The Ocean Heat Uptake Efficiency (OHUE) quantifies the ocean's ability to mitigate surface warming through deep heat sequestration. Despite its importance, the main controls on OHUE, as well as its nearly two-fold spread across contemporary climate models, remain unclear.
We argue that OHUE is primarily controlled by the strength of mid-latitude ventilation in the background climate, itself related to subtropical pycnocline depth and ocean stratification. This hypothesis is supported by a strong correlation between OHUE and pycnocline depth in the CMIP5 and CMIP6 under RCP85/SSP585, as well as in MITgcm. We explain these results through a regional OHUE decomposition, showing that the mid-latitudes largely account for both: (1) global heat uptake after increased radiative forcing and; (2) the correlation between pycnocline depth and OHUE. Coupled with the nearly equivalent inter-model spreads in OHUE/pycnocline depth, these results imply that mid-latitude ventilation also dominates the ensemble spread in OHUE. Our results provide a pathway towards observationally constraining OHUE, and thus future climate.
   
\end{abstract}


%
%

\section{Introduction}
\label{intro}

The ocean has absorbed around 90$\%$ of the excess energy in the climate system during the industrial age \cite{Cheng2017Improved2015,Meyssignac2019MeasuringImbalance}. This quantity is not invariant in time, nor across climate models, but is instead set by dynamical ocean processes. How efficiently these processes sequester heat away from the ocean surface partially determines the rate of surface warming. This study aims to better understand what controls this ocean heat sequestration and identify the dominant factor driving its spread across contemporary atmosphere-ocean climate models (AOGCMs).

In practice, the rate of heat uptake relative to a given surface warming is called the Ocean Heat Uptake Efficiency \cite <OHUE, >{Raper2002TheResponse,Gregory1997TheAdjustment}. OHUE is defined as the change in global-mean rate of ocean heat uptake (N, W m$^{-2}$) per unit change in global surface temperature, here represented  by the sea surface temperature (SST), since our study focuses on ocean processes (e.g., Gregory et al., in review):

\begin{equation}
\mathrm{OHUE} \equiv \frac{\mathrm{N}}{\mathrm{SST}}.
\label{e1}
\end{equation}

Over decadal to centennial timescales, the OHUE has around 30-75$\%$ the influence of the global radiative feedback on the rate of surface warming \cite [Gregory et al., in prep]{Kuhlbrodt2012OceanChange} and shares the units of a radiative feedback (Wm$^{-2}$K$^{-1}$). While OHUE is useful to quantify how ocean processes mitigate surface warming, it is also quite uncertain, ranging by a factor of one to two across AOGCMs \cite <e.g.,>[Gregory et al., in review, and Fig. \ref{f2} of this study] {Kuhlbrodt2012OceanChange}. 

The drivers of this spread are unclear, in part because the processes that govern OHUE are not fully understood. Past studies have linked the inter-model spread in OHUE to the background strength of the Atlantic Meridional Overturning Circulation, or AMOC \cite{Kostov2014ImpactChange, Romanou2017RoleTracers}, given a notable correlation between the background AMOC strength and depth and of heat and passive tracer transport in AOGCMs. An associated correlation between OHUE and the AMOC strength was also demonstrated across a number of different AOGCMs \cite{Winton2014a}, including those in CMIP5 and CMIP6 (Gregory et al., in prep.). Yet, the mechanism linking AMOC to OHUE is not obvious, given that relatively little anthropogenic heat uptake occurs in the North Atlantic \cite{Saenko2021ContributionAogcms}, as compared to wind-driven subduction regions in the Southern Ocean and mid-latitudes  \cite{Kuhlbrodt2012OceanChange,Frolicher2015DominanceModels,Armour2016,Shi2018EvolvingUptake,Zanna2019GlobalTransport,Newsom2020TheUptake,Cheng2022PastWarming}. 

More recent work instead argues that the correlation of AMOC and OHUE emerges because of their shared dependence on another aspect of the ocean state, for instance, the strength of transient ocean eddies, parameterized by $\kappa_{GM}$ \cite <>[Gregory et al., in prep] {Saenko2018ImpactModel}. Indeed, OHUE correlates negatively with $\kappa_{GM}$ across CMIP3 models \cite{Kuhlbrodt2012OceanChange}. Similarly, \citeA{Saenko2018ImpactModel} showed that although decreasing $\kappa_{GM}$ in an ocean model (NEMO3.4) increased both AMOC strength and OHUE, the latter change was associated with strengthened Southern Ocean ventilation, not AMOC changes. While these studies convincingly show the influence of $\kappa_{GM}$ on OHUE,  the wider relevance of this influence is hard to discern, given that many other processes affect both Southern Ocean and global heat uptake in addition to ocean eddies \cite<e.g.>{Exarchou2015OceanIntercomparison,Lyu2020ProcessesWarming,Morrison2022VentilationPycnocline}. For instance, recent work links higher Southern Ocean surface salinity to more efficient ocean heat uptake in CMIP6 models \cite{Liu2023TheSalinity}, consistent with a similar link between Southern Ocean surface salinity and global carbon uptake \cite{Terhaar2021SouthernSalinity}. Theory also points to the Southern Ocean mean-state wind strength as a key control on the depth of heat uptake under climate forcing \cite{Marshall2014AChange}.

Here we propose a more generalized perspective on OHUE, one not based on singular process such as the AMOC, transient ocean eddies, or Southern Ocean surface salinity or winds. We propose that OHUE will be primarily controlled by the depth and efficiency of ventilation from the mid-latitudes, itself set by many processes and reflected by the depth of the pycnocline (here meaning, the pycnocline between $\approx$ 60$^\circ$N/S), and the distribution of ocean stratification more broadly. 

Our reasoning is based on the idea that, as noted above, the majority of global anthropogenic heat and carbon uptake \cite <e.g.,>{Frolicher2015DominanceModels,Zanna2019GlobalTransport,Cheng2022PastWarming} occurs within the mid-latitudes, much of which enters the interior along sloping isopycnals \cite{Jackett1997,Church1991AExpansion, Saenko2021ContributionAogcms,Morrison2022VentilationPycnocline} akin to a passive tracer \cite{Winton2013ConnectingClimate,Gregory2016IntercomparisonOf,Couldrey2021WhatForcing}. Conceptually, a deeper subtropical pycnocline would enable isopycnals outcropping in the mid-latitudes to penetrate more deeply into the interior ocean, enabling deeper along-isopycnal heat subduction under anthropogenic forcing. This concept is depicted schematically in Fig. \ref{f1}, which illustrates the differences in mid-latitude heat penetration between a state with a deep pycnocline to one with a shallow pycnocline. We label these the ``High OHUE'' and  ``Low OHUE'' states, respectively. Critically, we propose that this connection between the background pycnocline depth and the depth of mid-latitude heat penetration under forcing is not a happenstance of ocean geometry. Instead, the mid-latitudes dominate pycnocline ventilation (between latitudes $< 60^\circ$N/S) \cite{Khatiwala2012VentilationAge,Sallee2010SouthernVentilation,Newsom2020TheUptake,Morrison2022VentilationPycnocline}. Thus the pycnocline depth should largely reflect how deeply these water masses penetrate the interior in the mean state,  
signifying the ocean volume available to sequester heat sourced in these regions under climate forcing. 

This argument is supported by the relationship between Southern Ocean ventilation and both pycnocline depth \cite{Gnanadesikan1999,Nikurashin2011b,Nikurashin2012b,TimeDependentResponseoftheOverturningCirculationandPycnoclineDepthtoSouthernOceanSurfaceWindStressChanges} and OHUE \cite{Marshall2014AChangeb,Saenko2018ImpactModel} implied by theory and found in Green's Function experiments \cite{Newsom2020TheUptake}. It is also supported by the relationship between the vertical density gradient in the Southern Ocean  \cite{Bourgeois2022Stratification55S,Terhaar2021SouthernSalinity,Liu2023TheSalinity} and the efficiency of regional and global heat and carbon uptake. Further support comes from recent work \cite{Newsom2022RelatingWarming} demonstrating a strong correlation between the global pattern of pynocline depth (defined as the e-folding depth of the vertical density profile) and of passive heat storage under radiative forcing. A similar relationship was noted for passive heat and carbon sequestration patterns \cite{Bronselaer2020HeatChanges}. Together, these studies imply that the same processes that establish the depth of the pycnocline --- in large part, ventilation from the Southern Ocean and mid-latitudes --- are key controls on the sequestration of heat and other tracers in the interior ocean. 

In what follows, we show that in both CMIP5-6 and the MITgcm: (1) there is a strong relationship between OHUE and the depth and stratification of the pycnocline depth; and (2) this relationship emerges because of the dominant role of mid-latitudes in setting OHUE. We will argue that differences in pycnocline depth (and, by implication, mid-latitude ventilation) across models also helps to explain inter-model spread in OHUE. We test these ideas by decomposing global OHUE into four regional components, namely the mid-latitudes, tropical and subtropical latitudes, northern high latitudes, and southern high latitudes, as depicted on Fig. \ref{f1}. This regional decomposition is used to quantify each region's relationship to global OHUE and pycnocline depth. The models and metrics used are described in Section \ref{methods} and our results are presented in Section \ref{results}. We discuss and summarize our findings in Section \ref{discussion}.

\section{Methods}
\label{methods}
\subsection{Model Ensembles} 

We primarily explore the link between OHUE and stratification in 28 fully-coupled CMIP5-6 models, supported by a set of perturbed parameter and surface forcing pattern experiments in the ocean-only MITgcm. The CMIP5 and CMIP6 models are subject to a historical and then an RCP 8.5 (CMIP5) or SSP585 (CMIP6) future forcing scenario and listed in Fig. \ref{f2}a. Anomalies in all CMIP variables will be defined as the average value of each over years 2090-2100 minus its average during 1850-1890. Note that the regression coefficients between various metrics and OHUE components (defined below) are not significantly different between CMIP5 and CMIP6 (at the 5\% level), thus we combine the ensembles unless otherwise noted.

The MITgcm ensemble, described in depth by \citeA{Huber2017DriversUptake,Zanna2019UncertaintyPredictions}, is used to support the CMIP5-6 results by testing the sensitivity of OHUE to different processes. Individual members vary in either their mesoscale eddy diffusivity  ($\kappa_{GM}$), vertical diffusivity ($\kappa_\nu$), or surface forcing pattern ($F$). After a 1000 year spin-up, all experiments are driven by surface salinity, temperature, and air-sea fluxes from the $1\%$CO$_2$ CMIP5 experiment, either in the multi-model mean (for $\kappa_{GM}$ and $\kappa_{\nu}$ perturbations), or from individual models (for $F$ perturbations). All anomalies are calculated at the time of CO$_2$ doubling. Ranges of OHUE for each set of experiments are shown in Fig. \ref{f2}b.

\subsection{OHUE definitions}

Traditionally, OHUE is defined as a global average quantity, e.g., Eq. \ref{e1}. Here, we introduce a complementary definition of a ``regional'' OHUE. This allows us to partition the global OHUE into several regional components, each OHUE$_{R}$:
\begin{equation}
    \mathrm{OHUE}_{R} \equiv \frac{\int_{A_R} \mathrm{N}(x,y) dA}{A_G\mathrm{SST}}.
    \label{e2}
\end{equation}

\hfill \break
Here,  $A_R$ is the surface area of a given region, $A_G = \sum_{R} A_R$ is the global surface area, $\mathrm{N}(x,y)$ is the local anomaly in surface heat flux at each latitude and longitude, $\mathrm{SST}$ is the global mean sea-surface temperature anomaly, and the units of $\mathrm{OHUE}_{R}$ are Wm$^{-2}$K$^{-1}$. In this study, we focus on four regional components associated with different mechanisms of heat uptake, OHUE$_{R}$=(OHUE$_{MidLat}$,OHUE$_{LowLat}$,OHUE$_{SHighLat}$, OHUE$_{NHighLat}$): 
\begin{enumerate}
    \item OHUE$_{MidLat}$: the mid-latitude OHUE calculated with Eq. \ref{e2} using the area between 30-60$^\circ$ in both hemispheres; 
    \item OHUE$_{LowLat}$: the subtropical and tropical OHUE, calculated between 30$^\circ$N/S; 
    \item OHUE$_{SHighLat}$: the Southern high-latitude OHUE, calculated south of 60$^\circ$; and 
    \item OHUE$_{NHighLat}$: the Northern high-latitude OHUE, calculated north of 60$^\circ$N. 
\end{enumerate}

Together, 
\begin{equation}
     \mathrm{OHUE} = \mathrm{OHUE}_{MidLat} +  \mathrm{OHUE}_{LowLat} +  \mathrm{OHUE}_{SHighLat} + \mathrm{OHUE}_{NHighLat},
\label{reg_comp}
\end{equation}

where OHUE is the global OHUE (Eq. \ref{e1}) and equivalently in Eq. \ref{e2} using global ocean area, A$_G$.

\subsection{Stratification Metrics} 
\label{metrics}

Our hypothesis centers around global stratification. To illustrate the interior stratification pattern, we use the zonal-mean (denoted by an overbar) Brunt--Väisälä frequency, $\overline{N^2} (y,z)= g/\rho_0 \partial \rho/\partial z$, assuming $\partial \rho/\partial z>0$, calculated from $\rho(x,y,z)$ and then zonally averaged. Here $\rho$ is the potential density referenced to 2000$dbar$, which we use to avoid biasing our density coordinate to the surface or abyssal ocean, especially important for Southern Ocean and intermediate water masses \cite<e.g.,>{Newsom2018ReassessingCirculation,Waugh2019ResponseWinds}. However, all results are robust to the choice of reference pressure. 

We characterize this stratification pattern through a representative scalar metric --- the pycnocline depth. While the pycnocline is often identified as the bottom of the shallow subtropical gyres (e.g., \citeA{Feucher2019SubtropicalOcean}),  our goal is to instead identify the depth to which mid-latitude sourced water masses penetrate. This depth is also co-located with a significant change in the vertical stratification (see Fig. S1), which we approximate in practice as the e-folding depth of the vertical density distribution, modified to exclude the strongly-stratified surface gyres as follows (also see the SI for expanded discussion).  We first take the zonal mean of the density field, $\overline{\rho}(y,z)$. We then mask out all density classes in $\overline{\rho}$ less than $\overline{\rho}_{gyre}$, where $\overline{\rho}_{gyre}$ is the zonal-mean density at the base of the mixed layer (350m, see \citeA{BuongiornoNardelli2017SouthernData}) at 45$^\circ$S i.e.,\ $\overline{\rho}_{gyre} \equiv \overline{\rho}(45S,350 m)$. This leaves the ``interior density field,'' $\overline{\rho^*} (y,z)$, where $y$ is latitude, $z$ depth, and the superscript ``$^*$" signifies all density classes greater than isopycnal $\overline{\rho}_{gyre}$, isolating the density profile of the waters predominantly sourced in the mid- and high latitudes. Since density increases downwards, $\overline{\rho^*}(y,B)-\overline{\rho^*}(y,z)>0$, where $z=B(y)$ at the bottom of the ocean.
Hence we derive a normalized vertical density coordinate $\overline{\rho}_{norm}$ which ranges from zero at the ocean bottom to unity at the shallowest depth considered:
\begin{equation}
 \overline{\rho}_{norm}(y,z) = \frac{\overline{\rho^*}(y,B)-\overline{\rho^*}(y,z)}{\mbox{max}_{z}\big(\overline{\rho^*}(y,B)-\overline{\rho^*}(y,z)\big)},
 \label{pyc}
\end{equation}
where ``$\mbox{max}_{z} X(y,z)$'' means the largest value of $X$ for given latitude $y$. The largest value of the denominator in Equation~\ref{pyc} belongs to the shallowest $z$ considered, either the surface or the depth of $\rho_{gyre}$. Thus $\overline{\rho}_{norm}$ captures the shape of vertical variation in zonal-mean density, below the shallow surface gyres, across models. We define the pycnocline depth, $d$, as the depth at which $\overline{\rho}_{norm}=1/e$ at each latitude (see Fig. S1), which assumes $\overline{\rho}_{norm}$ can be approximated as an exponential profile, such that $\overline{\rho}_{norm} = e^{-z/d}$. In what follows, we will refer to the average pycnocline depth between 60$^\circ$S/N as the "pycnocline depth" unless otherwise noted. 

Notably, the pycnocline depth covaries strongly with other measures of mid-latitude ventilation strength, such as the slopes or stratification of Southern Ocean isopycnals (at $R=0.9$ and $R= 0.95$, respectively), as discussed in the SI.  This interconnection, between Southern Ocean and global stratification, affirms the dominance of the Southern Ocean in ventilating the global pycnocline, e.g.,  \citeA{Sallee2010SouthernVentilation,Khatiwala2012VentilationAge,Morrison2022VentilationPycnocline}.

\section{Results}
\label{results}
\subsection{Global heat uptake efficiency}
\label{resultsa}

We begin by examining global aspects of OHUE ($=\mathrm{N}/\mathrm{SST}$) in CMIP5-6. As noted in Section \ref{intro}, there is a wide spread in OHUE across CMIP5-6 models. The ratio of the standard deviation in OHUE to its ensemble mean, i.e. the spread, is 17$\%$ in CMIP5 and 13$\%$ in CMIP6 (Fig. \ref{f2}a). Individually, the spreads in $\mathrm{N}$ is 15$\%$ and 16$\%$ in CMIP5 and CMIP6, and 21$\%$ and 22$\%$ in $\mathrm{SST}$. The spread in OHUE is generally smaller than N or $\mathrm{SST}$ individually because N and $\mathrm{SST}$ are correlated (Gregory et al., in review), more so in CMIP6, in which the correlation between $\mathrm{N}$ and $\mathrm{SST}$ is stronger (at $R\approx0.88$) than in CMIP5 ($R\approx0.68$). The mean OHUE is also smaller in CMIP6, which may result from the higher climate sensitivity in this ensemble \cite{Zelinka2020CausesModels}.

Differences in OHUE across CMIP5-6 models are associated with different vertical profiles of warming, quantified by the heat storage per unit depth and latitude, ${H}(y,z) \equiv \int_{0}^{360} \theta(x,y,z) dx$ (units K~ m), where $\theta$ is the anomaly in ocean temperature (see Section \ref{methods}). Models also differ in their total heat storage, $\mathcal{H} =\int_B^0\int_{-90}^{90}Hdydz$.  Therefore, to compare the pattern of heat storage in each model, relative to one another, we consider the normalized heat storage pattern $H_{n}(y,z) \equiv H(y,z)/\mathcal{H}$, similar to Gregory et al., in review. Fig. \ref{f2}b-c illustrates the correlation of OHUE and  $H_{n}$ with latitude and depth, the latter for which we sum $H_{n}$ meridionally. Figs.  \ref{f2}b-c show that greater OHUE is associated with relatively more heat storage below $\approx$600m, and less heat storage above $\approx$600m  (Fig.   \ref{f2}c), particularly within the Southern mid-latitudes, $\approx$ 55S-30S, see Fig.~\ref{f2}b. This pattern implies a larger net global heat flux across $\approx600$m in models with higher OHUE, consistent with \citeA {Saenko2018ImpactModel,Kostov2014ImpactChange}; Gregory et al., in review. 

Critically, while the correlation of OHUE and $H_{n}$ is positive between $\approx$600-3000m depth (Fig. \ref{f2}b), the majority ($>80\%$) of anomalous heat is stored in the upper 2000 meters in all models. Thus, inter-model differences in vertical heat storage patterns, and the correlation of these patterns to OHUE, are most impactful in the upper 2000m, i.e., they involve larger quantities of heat. To illustrate this, we also show the regression coefficient for OHUE on $H_{n}$ at each latitude  (Fig. \ref{f2}d) and depth (Fig. \ref{f2}e). This regression pattern also crosses zero at $\approx$ 600m, but peaks at $\approx$ 1200m --- this slope will be larger, for a given correlation strength,  where the mean heat content across models is greater. Regression patterns illustrate that higher OHUE  is primarily associated with processes moving heat from the surface ocean ($<$600 $m$ depth) into intermediate depths ($\approx$600-2000 $m$).

We hypothesize that greater OHUE, and thus greater heat storage across intermediate depths, is linked to a deeper global pycnocline and, correspondingly, weaker pycnocline stratification in the background state. To test this hypothesis, we examine the relationship between OHUE and the zonal-mean stratification, $\overline{N^2}$ (defined in Section~\ref{metrics} and Fig. \ref{f3}a for the ensemble mean) in CMIP5-6. Fig. \ref{f3}b shows the correlation of these quantities as a function of latitude and depth. A general pattern emerges in Fig. \ref{f3}b, in which greater OHUE is associated with a weaker stratification of the water masses that outcrop at the mid-latitude surface (latitudes $\approx$60$^\circ$-30$^\circ$ in both hemispheres, see isopycnals in Fig. \ref{f3}a ) and fill the basin interior above $\approx$1300$m$. 

The low-stratification signature in Fig. \ref{f3}b is  largely bounded below by the pycnocline depth (defined in Section~\ref{metrics}), which correlates significantly with OHUE between $\approx$60S-N. This is evident in  Fig. \ref{f3}b, which shows the pycnocline depth across in CMIP5-6 models colored by OHUE strength. It is quantified by the correlation of OHUE with the pycnocline depth of $R=0.83$ in CMIP5-6, meaning a deeper pycnocline corresponds to greater OHUE, Fig. \ref{f3}c. In the MITgcm, the correlation with pycnocline depth even stronger, at 0.92.  Together, these clear and consistent relationships support our hypothesis that OHUE is closely related to the stratification of the global pycnocline.

\subsection{Regional heat uptake efficiency}
\subsubsection{Mid-latitudes}
\label{resultsb}

Our regional OHUE decomposition (Eq. \ref{reg_comp}) clarifies the importance of the mid-latitude regions in setting the global relationships between OHUE and stratification discussed in Sec. \ref{resultsa}. Component OHUE$_{MidLat}$ encapsulates the efficiency of heat uptake, relative to global mean surface temperature, from the region between 30-60$^\circ$ N/S. This region accounts for 70$\%$ of the total global anomaly in ocean heat uptake (Fig. \ref{f1}b) in CMIP5-6 during the period considered (2090-2100), as well as around 70$\%$ of the variance in OHUE, in both CMIP5 and CMIP6 (Fig. S4). This is consistent with the historical dominance of observed heat uptake from these regions \cite<e.g.,>{Frolicher2015DominanceModels,Zanna2019GlobalTransport,Cheng2022PastWarming}. 

To probe the relationship of  OHUE$_{MidLat}$ and $\overline{N^2}$, we again calculate their correlation at each latitude and depth (Fig. \ref{f4}a). As before, a clear fingerprint emerges, linking greater OHUE$_{MidLat}$ to more weakly stratified water-masses above the pycnocline, both within the Southern Ocean and in the basins to its north. While the relationship of OHUE$_{MidLat}$ to the stratification is qualitatively similar to that of OHUE (Fig. \ref{f3}b), it is stronger both within the Southern Ocean, where local stratification and OHUE$_{MidLat}$ are correlated at $R>=0.75$ in CMIP5-6 (Fig. S5), and more coherent into the interior across depths $\approx$ 500-1500$m$ (Fig. \ref{f4}a).  The relationship between OHUE$_{MidLat}$ and pycnocline depth is similarly robust,  This is apparent in Fig. \ref{f4}a, which depicts the pycnocline depth across CMIP5-6 models colored by the strength of OHUE$_{MidLat}$, as well as the stronger correlation between the OHUE$_{MidLat}$ and pycnocline depth, at $R=0.86$, Fig. \ref{f4}b. In MITgcm, pycnocline depth and OHUE$_{MidLat}$ are correlated at $0.87$, which, while slightly weaker than for OHUE, is remarkably consistent with CMIP5-6. The similarity in the strength of these relationships across ensembles corroborates the idea that more heat can be absorbed in the mid-latitudes, for a given global surface warming, when regional ventilation is strong, as evidenced by a deeper, less stratified pycnocline in the background state. 



\subsubsection{The Low and High Latitudes}
\label{resultsc}
OHUE is also substantially influenced by processes outside of the mid-latitudes. The remaining $\approx$30$\%$ of OHUE in CMIP5-6 (as regionally partitioned in Section \ref{methods}) is accounted for by the southern high latitudes (OHUE$_{SHighLat}\approx$13$\%$), the northern high latitudes  (OHUE$_{NHighLat}$ $\approx$9$\%$) and the low-latitudes (OHUE$_{LowLat}$ $\approx$10$\%$)  (Fig. \ref{fs6}c) on average.  To understand how these regions influence the relationship of OHUE and the pycnocline depth, we again examine the point-wise correlation between $\overline{N^2}$ and each component (Fig. S5). Unlike OHUE and OHUE$_{MidLat}$, no clear or physically meaningful pattern emerges in this calculation for any of these regional components, with the exception of OHUE$_{SHighLat}$, which is higher for models with weak full-depth stratification south of 60S (Fig \ref{fs6}). Accordingly, we find no clear relationship between each component  --- OHUE$_{SHighLat}$,  OHUE$_{NHighLat}$, and OHUE$_{LowLat}$ --- and pycnocline depth in CMIP5-6. Essentially, the high and low latitude regions  serve as ``noise'' in the relationship between OHUE and stratification, as quantified here, reinforcing that this relationship is mediated through mid-latitude processes.

\section{Discussion and Conclusions}
\label{discussion}

Our results reveal a strong connection between global OHUE and global stratification, as quantified by the pycnocline depth. 
We argue that the connection exists because the pycnocline is a proxy for the depth of mid-latitude ventilation in the background state, and that these same ventilation processes make the largest contribution to global OHUE under anthropogenic forcing. A corollary of these results is that heat uptake efficiency outside the mid-latitudes has little direct connection to subtropical pycnocline depth, perhaps unsurprisingly, as no clear mechanism would predict such a connection. 

\begin{figure}[h]
    \centering
\includegraphics[width=\textwidth]{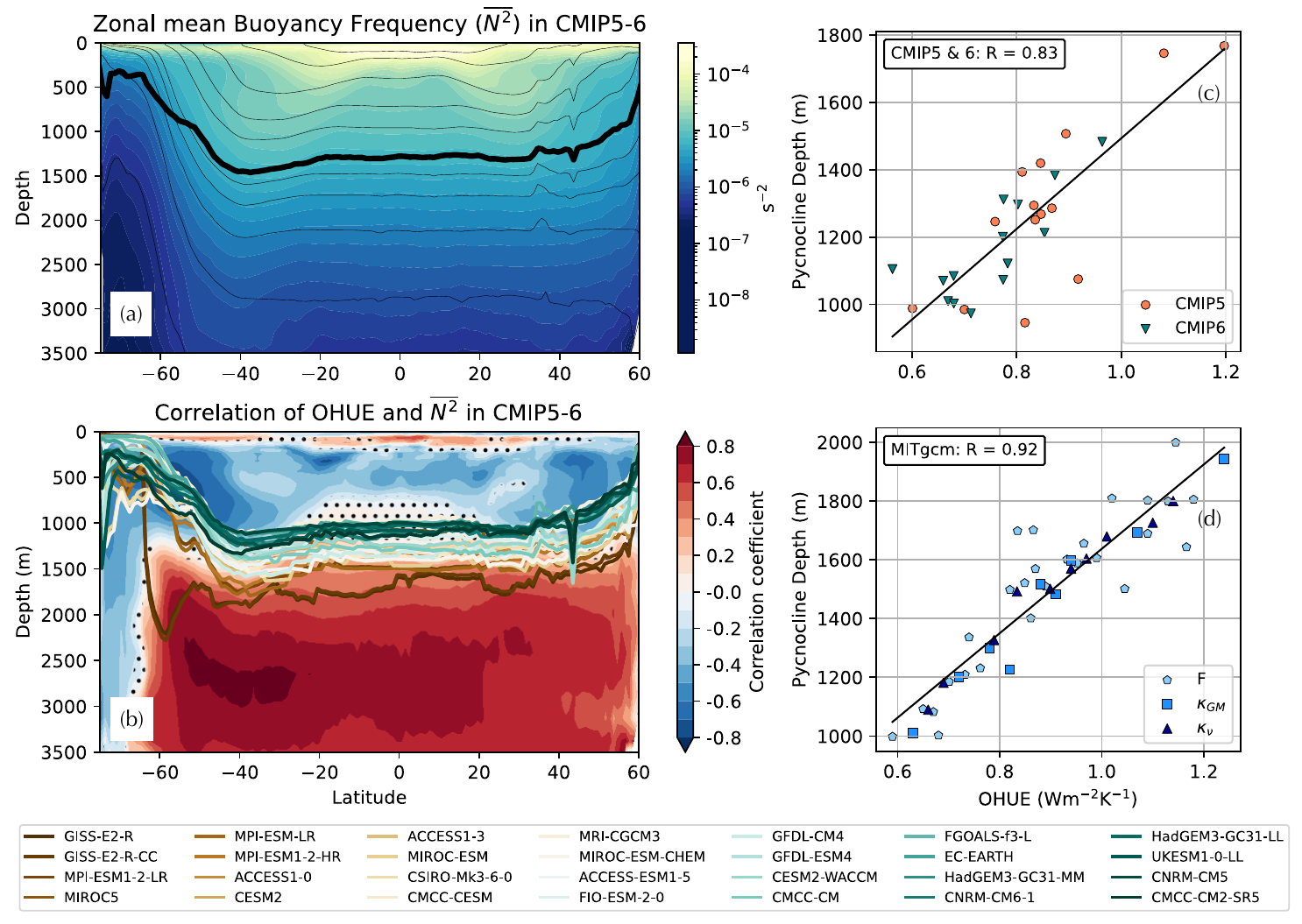}\caption{a) The CMIP5-6 ensemble mean zonal-mean buoyancy frequency ($\overline{N^2}$) is shown to illustrate mean stratification distribution with depth and latitude. Overlaid on (a) is the ensemble mean pycnocline depth (thick black) and  several isopycnals (spaced by $\approx0.5  kg/m^3$  above the pycnocline and $\approx0.1  kg/m^3$ below it) to show where waters above $\approx$ 1500 $m$ outcrop. b) Point-wise correlation between $\overline{N^2}$ and OHUE in CMIP5-6 (b). Overlaid on (b)  is pycnocline depth, defined via Eq. \ref{pyc}, and colored from high to low OHUE across models (brown to green colors). c-d) OHUE versus the average pycnocline depth in CMIP5-6 (c) and the MITgcm (d) . Note that on (b), the stippling shows where the correlation is not significant to the 95th percentile. Also note that $\overline{N^2}$ is 2-3 orders of magnitude smaller below $\approx$ 1500  m than above it, so while these deeper correlations are significant, they involve small vertical density differences. This distinction is evident in the regression of OHUE on $\overline{N^2}$ (Fig. S5).  }
    \label{f3}
\end{figure}

\begin{figure}[h]
    \centering
\includegraphics[width=\textwidth]{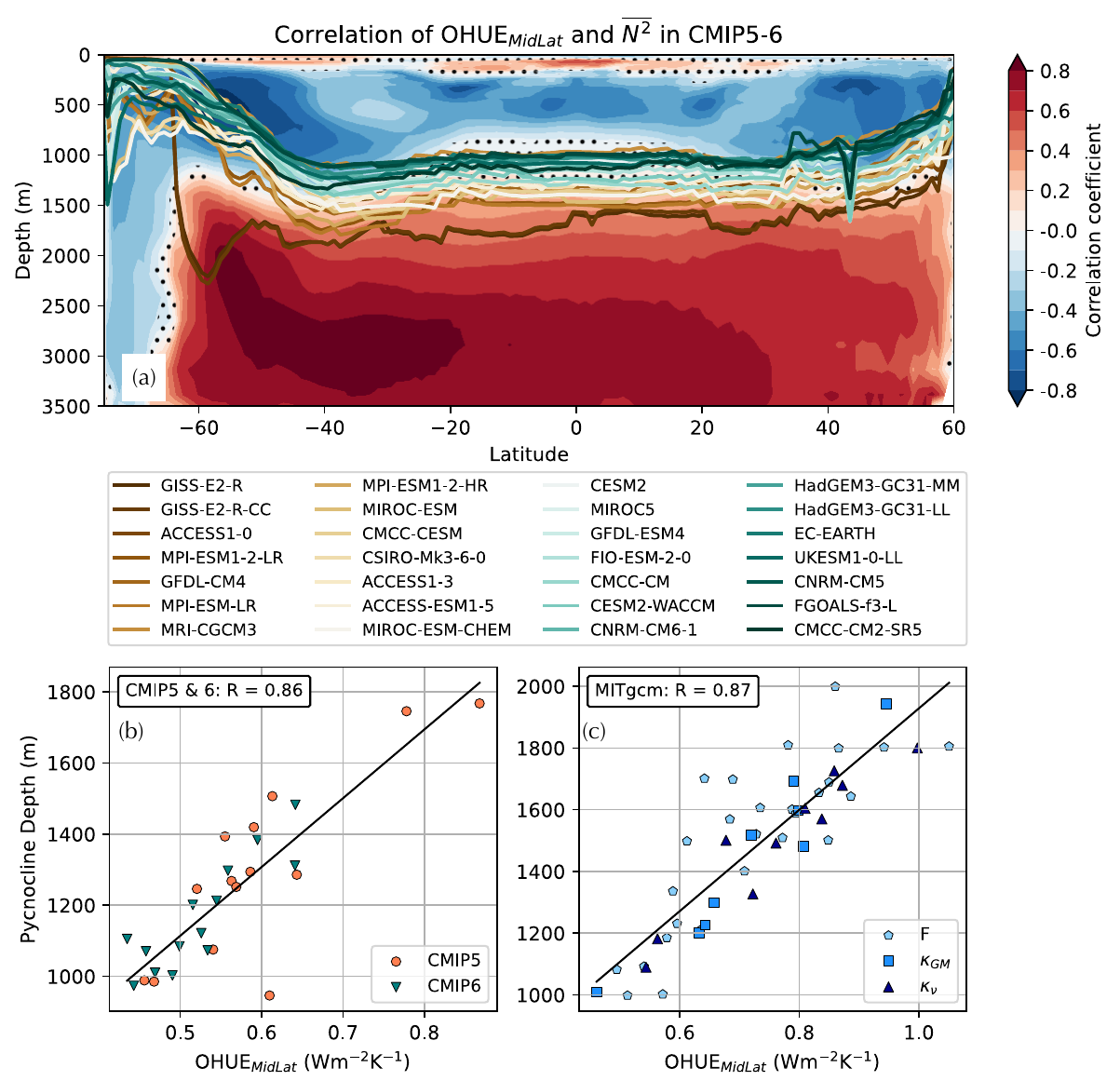}\caption{a) The point-wise correlation between OHUE$_{MidLat}$ and $\overline{N^2}$  in CMIP5-6. Overlaid is the pycnocline depth, which is colored from high to low OHUE$_{MidLat}$. b) As in Fig. \ref{f3}c, but for the OHUE$_{MidLat}$ in CMIP5-6. c) As in Fig. \ref{f3}d, but for the OHUE$_{MidLat}$ in MITgcm. }
    \label{f4}
\end{figure}

These findings align with previous work relating both background stratification to Southern Ocean processes \cite{Gnanadesikan1999,Nikurashin2011b,Nikurashin2012b,Marshall2014AChange,TimeDependentResponseoftheOverturningCirculationandPycnoclineDepthtoSouthernOceanSurfaceWindStressChanges}, and the depth of heat penetration to the vertical density profile (\citeA{Marshall2014AChange}). Studies that connect high OHUE to weaker Southern Ocean eddy activity (i.e., lower $\kappa_{GM}$, \citeA[Gregory et al., (in review)] {Kuhlbrodt2012OceanChange,Saenko2018ImpactModel} are particularly relevant, since, all else being equal, reducing $\kappa_{GM}$ will increase Southern Ocean ventilation and pycnocline depth (Figs. \ref{f3}-\ref{f4}, \citeA{Marshall2014AChange,Gnanadesikan1999}). However, ventilation is influenced by many additional processes, including Southern Ocean wind stress, and surface salinity, temperature, interior mixing, and surface buoyancy flux   \cite{Morrison2022VentilationPycnocline,Sallee2010SouthernVentilation,TimeDependentResponseoftheOverturningCirculationandPycnoclineDepthtoSouthernOceanSurfaceWindStressChanges}, all of which may influence OHUE. Indeed, Southern Ocean surface salinity correlates significantly with both OHUE \cite{Liu2023TheSalinity} and global anthropogenic carbon uptake in CMIP6 \cite{Terhaar2021SouthernSalinity}, consistent with the correlation between Southern Ocean vertical stratification and regional heat and carbon uptake efficiency \cite{Bourgeois2022Stratification55S}.  

Yet, our MITgcm results, in which individual parameters vary widely  --- including $\kappa_{GM}$, parameterized mixing, and surface salinity  --- show that no individual process controls OHUE. Instead, they effect OHUE through mid-latitude ventilation, as measured in the aggregate by pycnocline depth. Consistent with \citeA{Saenko2018ImpactModel}, there is a strong relationship between $\kappa_{GM}$ and OHUE, where stronger parameterized eddies lead to lower OHUE, a more stratified ocean and a shallower pycnocline (pycnocline depth and OHUE correlate at $R=0.98$ in the $\kappa_{GM}$ ensemble). The inverse is true for the $\kappa_{\nu}$ ensemble: greater $\kappa_{\nu}$  increases OHUE, stratification, and pycnocline depth (which also correlates with OHUE at $R=0.98$ ). The range of OHUE in the air-sea flux experiments ($F$) is less straightforward and evidences the strong sensitivity in the MITgcm to high-latitude, and north-south gradients of surface forcing \cite{Kostov2019AMOCMechanisms}. An important caveatto these strong correlations is that this model set up is coarse and the parameterizations are relatively simple \cite{Huber2017DriversUptake}, in an effort to isolate the relevant physics and underlying relationships. 

\begin{figure}[h]
    \centering
\includegraphics[width=\textwidth]{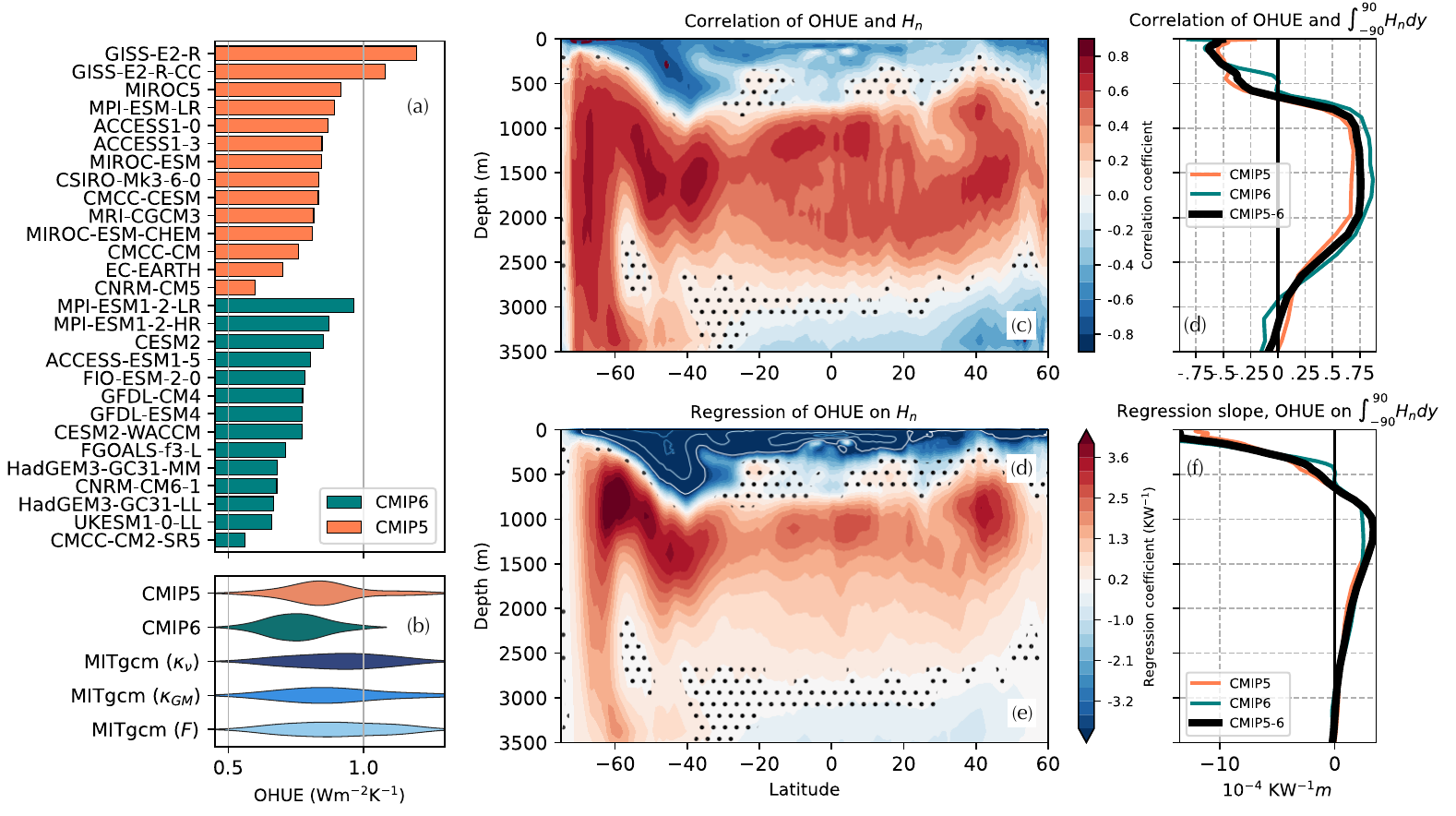}
\caption{a) OHUE (Eq. \ref{e1}) for CMIP5 models (orange) and CMIP6 models (teal). b) The distribution of OHUE in CMIP5, CMIP6 and in the MITgcm experiments. c) Correlation of OHUE and normalized global heat storage and $H_{n}(y,z)$  (see text). d) Same as (c), but for the meridional sum of $H_{n}$ at each depth. e) The regression coefficient of OHUE on $H_{n}(y,z)$. f) Same as (e), but for the meridional sum of $H_{n}$. Panels (e-f) illustrate that a key inter-model difference is the increased redistribution of anomalous heat from the upper ocean ($<\approx$600 m) into the intermediate ocean ($\approx$ 600-1500 m) in high OHUE models, as compared to low OHUE models. Note that stippling in this and the following figures shows where significance is below the 95th percentile.}
    \label{f2}
\end{figure}

The correlation of pycnocline depth and OHUE (Fig. \ref{f3}c) also implies that the spread in pycnocline depth in CMIP5-6 explains around 69$\%$ of the variance in OHUE (calculated here as $R^2$), and 74$\%$ of the spread in OHUE$_{MidLat}$. This is supported by a key difference between high and low OHUE CMIP5-6 models --- high OHUE models store relatively less heat in the upper $\approx$ 600 $m$ and more across intermediate depths (600-1500 $m$, Fig. \ref{f2}), consistent with \citeA{Kostov2014ImpactChange,Saenko2018ImpactModel} Gregory et al. (in review), and \citeA{Liu2023TheSalinity} . These depths are collocated with the clearest and most predictive differences in stratification between high and low OHUE models (see the strong negative correlations above $\approx$1500m in Figs. \ref{f3}b and \ref{f4}a), implying that the capacity to sequester heat here is linked to the weak stratification signature originating at the mid-latitude surface. Importantly, our MITgcm experiments highlight that better eddy closures in coarse resolution models \cite{Jansen2019TowardEddies,Zanna2020Data-DrivenClosures} would help to reduce this persistent spread in both global stratification and OHUE. 


The ocean physics we discuss in the context of OHUE may additionally modulate future climate through a surface warming ``pattern effect'' \cite{Armour2013b,Xie2020OceanChange,Gregory2016VariationPeriod,Stevens2016ProspectsSensitivity}. Recent work highlights how surface warming patterns are disproportionately influenced by heat uptake and warming in the Southern Ocean \cite{Lin2021TheCESM,Dong2022AntarcticEffect}, a region with a strong influence on climate sensitivity \cite{Andrews2015TheModels,Zelinka2020CausesModels}. However, the link between OHUE and the pattern effect is thus far unclear. Given the key role of the Southern Ocean and mid-latitudes in our study, a key goal of future work will be to develop a more unified understanding of how these regions influence transient climate change, incorporating their inter-dependent effects.

\begin{figure}[h]
    \centering
\includegraphics[width=\textwidth]{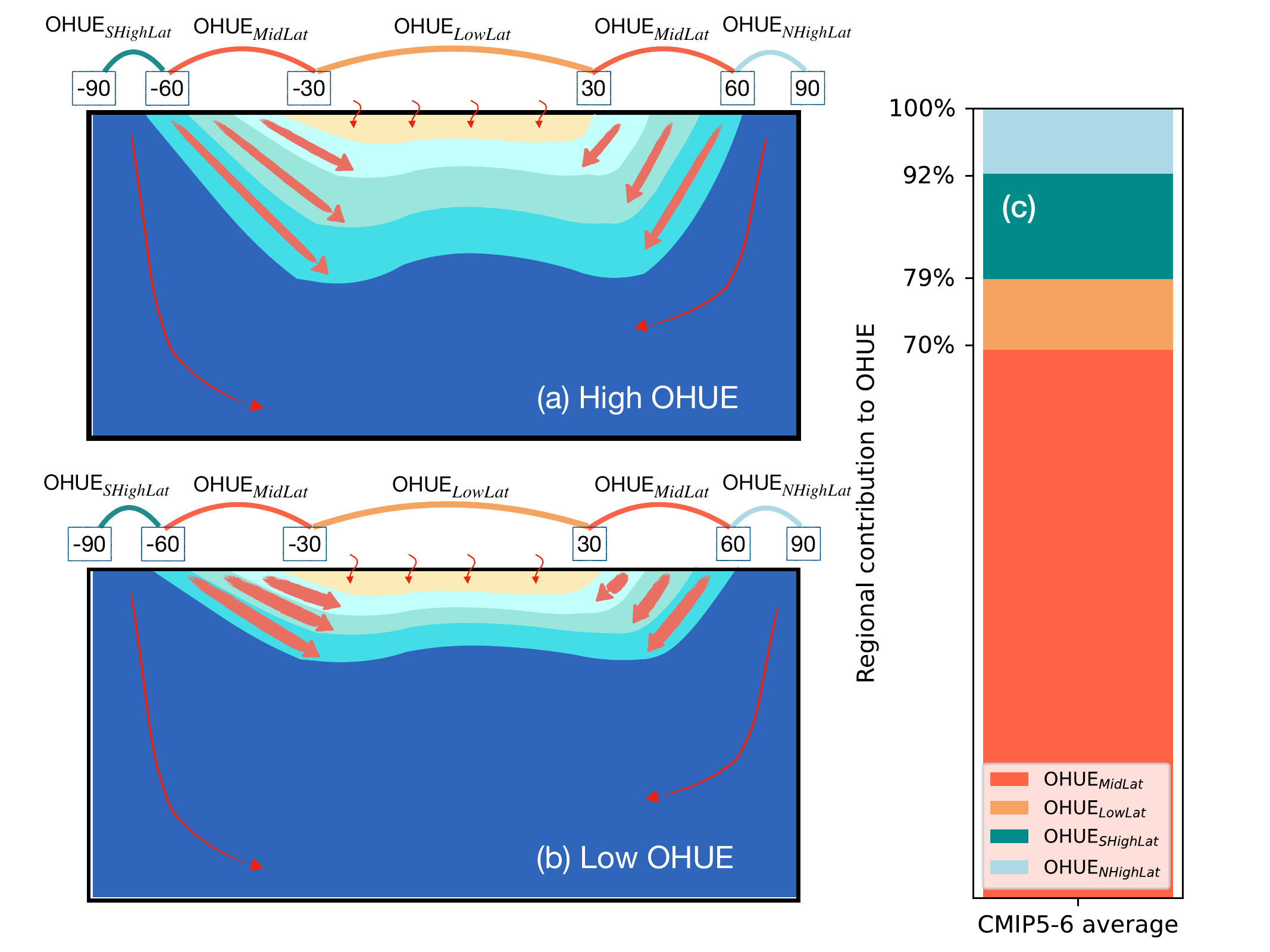}\caption{Left column) schematic of the relationship of OHUE and pycnocline depth, contrasting a high OHUE state (a) and a low OHUE state (b). These panels illustrate that a deeper pycnocline layer will be associated with weaker stratification within, and steeper outcropping slopes of, pycnocline water masses. All heat uptake that occurs along outcropping mid-latitude isopycnals will thus penetrate deeper into the interior for the deep pycnocline state (compare large red arrows between panels a and b) and more efficiently mitigate global surface warming. Here we depict the same net heat uptake (reflected by the size of the arrows), such that (a-b) differ only in the depth of heat penetration. Panels a-b also schematize the different regional components of global OHUE and their meridional extent (as described in Section \ref{methods}), while panel (c) shows the relative contribution of each regional component to global OHUE (or, OHUE), as discussed in Section \ref{results}. Note the size of the red arrows on a-b signifies the regional fraction of global ocean heat uptake. }
    \label{f1}
\end{figure}

To that end, a hope of this work is to constrain contemporary and future OHUE and warming patterns through oceanic observations, for instance, of pycnocline depth or stratification. Since the mid-latitudes account for around 70$\%$ of OHUE (Fig. \ref{f1}), identifying physical controls on OHUE outside the mid-latitudes is an important step for leveraging the relationships discussed here. Nonetheless, our results provide a promising path towards observationally narrowing the modeled range of OHUE, which may meaningfully reduce uncertainty in global sea level projections, surface warming, and potentially the ocean's long-term capacity store anthropogenic carbon, e.g., \citeA{Bronselaer2020HeatChanges,Terhaar2021SouthernSalinity,Bourgeois2022Stratification55S}.

\acknowledgments
The authors thank the WCRP's Working Group on Coupled Modelling, which is responsible for CMIP. ERN and LZ received support from the NSF OCE grant 2048576 on Collaborative Research: Transient response of regional sea level to Antarctic ice shelf fluxes and M$^2$LInES research funding by the generosity of Eric and Wendy Schmidt by recommendation of the Schmidt Futures program. JMG was supported by the European Research Council (ERC) under the European Union's Horizon 2020 research and innovation programme (grant agreement No 786427, project ``Couplet'').

%


%
%
%
%




%
%
%



\bibliography{references15}

%
%
%
%
%

\end{document}


%
%


\title{Supporting Information for ``Global Pycnocline depth constrains Ocean Heat Uptake Efficiency"}
%
%

%
%



\authors{Emily Newsom\affil{1}, Laure Zanna\affil{1}, Jonathan Gregory\affil{2}\affil{3}}

 \affiliation{1}{Courant Institute of Mathematical Sciences, New York University}
\affiliation{2}{National Centre for Atmospheric Science, University of Reading, Reading, UK}
 \affiliation{3}{Met Office Hadley Centre, Exeter, UK} 


%
%

%

\begin{article}

%
%

\noindent\textbf{Contents of this file}
\begin{enumerate}
\item Introduction
\item Text S1 to S2
\item Figures S1 to S6

\end{enumerate}

\noindent\textbf{Introduction}
To support the arguments in the main text, we include more information regarding our pycnocline depth metric and several additional figures, as described below.


\clearpage

\begin{figure}
\noindent\includegraphics[width=\textwidth]{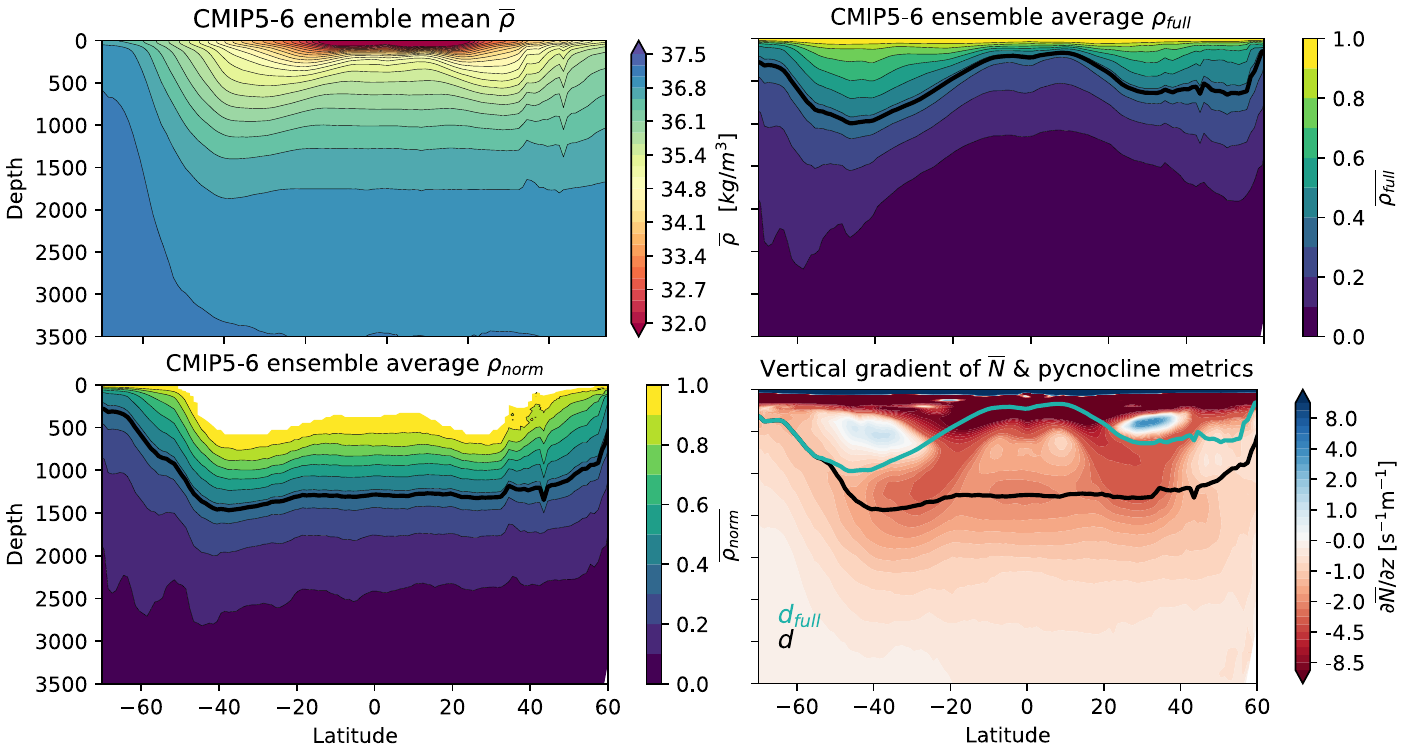}
\caption{a) Zonal-averaged density ($\overline{\rho}$) with depth in the CMIP5-6 ensemble mean, to show spacing of isopycnals. b) $\rho_{norm}$ (Eq. S1 and 4 in the main text), as calculated by removing the density classes comprising the subtropical and tropical gyres. Overlaid in black is $d=\rho(1/e)$, our definition of pycnocline depth. c) Same, but for $\rho_{full}$ (Eq. S2) and $d_{full}$. d) $\partial^2\overline{N} /\partial z^2$ (filled contours) overlaid with $d$ black) and $d_{fyll}$ (teal). Note $d$ generally falls at a local peak in stratification change (red colors) traceable that outcrops in the mid-latitude surface, particularly in the Southern Hemisphere.}
\label{fs1}
\end{figure}

\noindent\textbf{Text S1.}
As described in Section 2.3, our modified density variable, $\overline{\rho}_{norm}$, is calculated by from a modified density field $\overline{\rho^*}$, which excludes all densities less than all density classes in the subtropical surface gyres, water masses of density $\overline{\rho}_{gyre} \equiv \overline{\rho}(45S,350 m)$ or less. We then normalize  $\overline{\rho^*}$ so that it varies from $0$ to $1$ from the bottom, $z=B$, to the top-most grid  point, which is located at the surface of at the base of $\overline{\rho}_{gyre}$,  i.e.,

\begin{equation}
 \overline{\rho}_{norm}(y,z) = \frac{\overline{\rho^*}(y,B)-\overline{\rho^*}(y,z)}{\mbox{max}_{z}\big(\overline{\rho^*}(y,B)-\overline{\rho^*}(y,z)\big)},
 \label{pyc}
\end{equation}
or Eq. 4 in the main text.  To clarify the usefulness of excluding the surface gyres, we compare $\overline{\rho}_{norm}(y,z)$ calculated as in Eq. \ref{pyc} versus without omitting surface gyres:

\begin{equation}
 \overline{\rho}_{full}(y,z) = \frac{\overline{\rho}(y,B)-\overline{\rho}(y,z)}{\mbox{max}_{z}\big(\overline{\rho}(y,B)-\overline{\rho}(y,z)\big)},
 \label{pyc}
\end{equation}
in Fig. \ref{fs1}.  For reference, Fig. \ref{fs1}a shows $\overline{\rho}$, in the CMIP5-6 ensemble. Panels b-c compare $\overline{\rho}_{norm}$ to $\overline{\rho}_{full}$ , including the pycnocline depth defined as $d = \overline{\rho}_{norm}(1/e)$ and $d_{full} = \overline{\rho}_{full}(1/e)$, respectively. Fig. \ref{fs1}d  illustrates the vertical gradient of $\overline{N}(y,z)$, where $\overline{N} \propto \sqrt{\partial^2\rho/\partial z^2}$ and thus captures where the stratification in density changes rapidly. Note that here we use $\overline{N}$ instead of $\overline{N}^2$ to reduced the saturation of the figure. Overlaid on panel (d) are pycnocline metric $d$ and full depth metric, $d_{full}$. Panel (d)  illustrates that $d$ is generally co-located with an extreme in $\partial^2\overline{N} /\partial z^2$ $-$ a rapid decrease in stratification. Metric $d_{full}$, instead, shoals significantly at lower latitudes where watermasses $\approx>\rho_{gyre}$ are highly stratified and more closely corresponds to pycnocline with bounds the subtropical gyres, as opposed to the more deeply-penetrating watermasses of the mid-latitudes. 

\noindent\textbf{Text S2.}

To verify that our pycnocline metric (Section 2.3 in the main text) is a good representation of mid-latitude ventilation and stratification, we calculate two other scalar metrics: the zonal-average isopycnal slope and $\overline{N^2}$  in the Southern Ocean. Isopycnal slopes are defined as  

\[\overline{S}(y,z) = \frac{{\partial \overline{\rho}(y,z)}/{\partial z}}{ {\partial \overline{\rho}(y,z)}/{\partial y}}. \]
To obtain these scalar metrics, we average $\overline{N^2}$ and $\overline{S}$  between 450-900 meters and 57-47$^\circ$S, a region chosen to fall roughly above our pycnocline metric and below the winter mixed layer. We call these metrics the ``Southern Ocean stratification'' and ``Southern Ocean isopycnal slope,'' respectively.

\begin{figure}
\noindent\includegraphics[width=\textwidth]{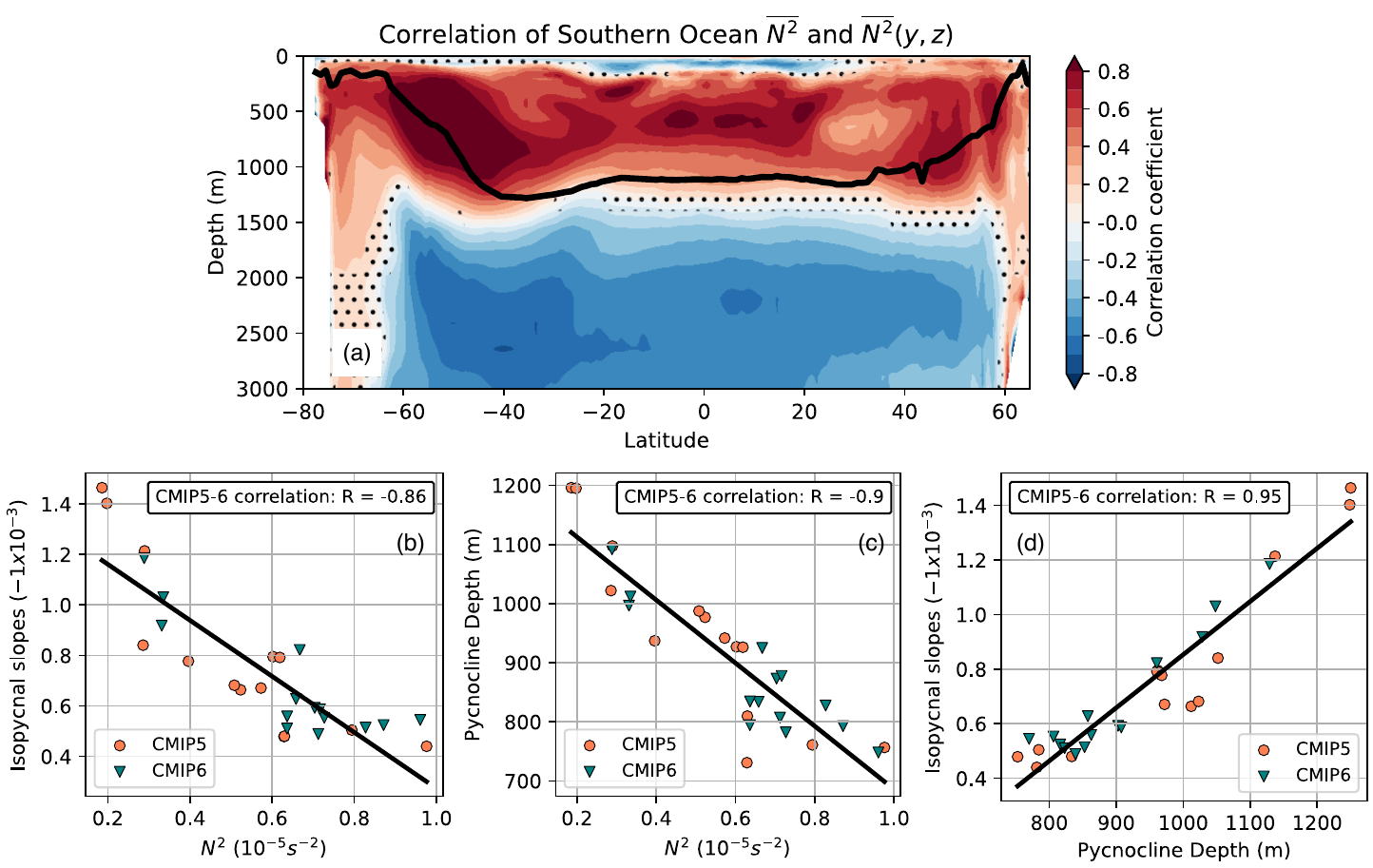}
\caption{a) The pointwise correlation between $\overline{N^2}$ at each latitude and depth for the
Southern Ocean stratification ( $\overline{N^2}$ averaged between 450-950 meters, latitudes 57◦S-47◦S) in
CMIP5-6. As schematized in Fig. 1 in the main text, weaker Southern Ocean stratification tends
to hold throughout the northern basins. The correlation between Southern Ocean stratification
and (a) Southern Ocean isopycnal slope and (b) subtropical pycnocline depth (averaged between
40◦N/S );. d) The correlation of subtropical pycnocline depth and Southern Ocean isopycnal slope.}
\label{fs2}
\end{figure}

As noted in the main text, these three metrics (including here the pycnocline depth) covary, since they measure related characteristics of the ocean state, interpreted as the strength of mid-latitude ventilation in the background state \cite{Gnanadesikan1999,Nikurashin2011b,Nikurashin2012b,TimeDependentResponseoftheOverturningCirculationandPycnoclineDepthtoSouthernOceanSurfaceWindStressChanges}. The correspondence between metrics bears out in CMIP5-6: Southern Ocean stratification is strongly correlated with $\overline{N^2}$ across similar depths north of 40$^\circ$S (R$\approx0.7-0.9$,  Fig. \ref{fs2}), as well as with global pycnocline depth (R$=0.9$). Similarly, Southern Ocean isopycnal slopes are also tightly coupled to  Southern Ocean $\overline{N^2}$ (R$=0.86$) and to pycnocline depth (R$=0.95$). This interconnection between Southern Ocean and global stratification affirms the dominance of the Southern Ocean in ventilating the global pycnocline, e.g.,  \citeA{Sallee2010SouthernVentilation,Khatiwala2012VentilationAge,Morrison2022VentilationPycnocline}.

\begin{figure}
\noindent\includegraphics[width=.8\textwidth]{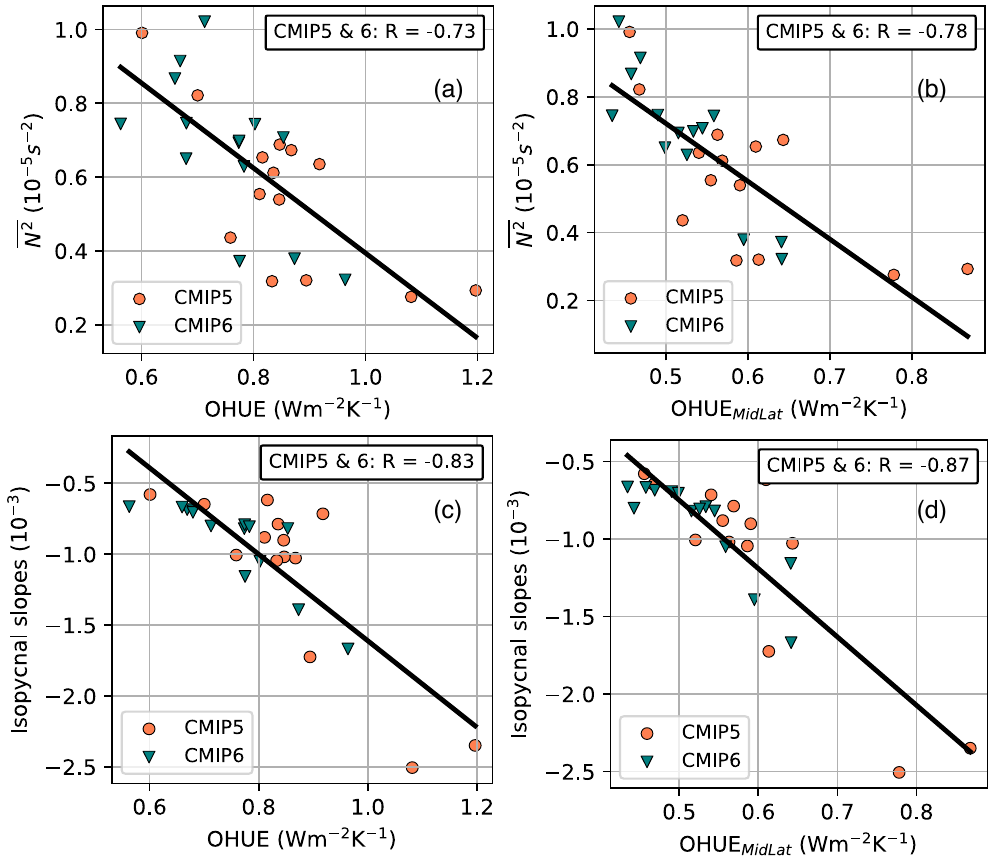}
\caption{The correlations of metrics of Southern Ocean stratification and OHUE and OHUE$_{MidLat}$. a-b) The correlation of Southern Ocean $\overline{N^2}$ and OHUE (a) and OHUE$_{MidLat}$ (b). c-d) same as (a-b), but for Southern Ocean isopycnal slope, $\overline{S}$}.
\label{fs5}
\end{figure}

Note that these Southern Ocean metrics ($\overline{N^2}$ and $\overline{S}$) also correlate strongly with OHUE and OHUE$_{MidLat}$ (Fig. \ref{fs5}). The co-variation of these metrics with pycnocline depth, and OHUE, support the interpretation of the pycnocline depth as a key measure of global stratification more broadly, and that stratification is strongly linked to OHUE.

 \bibliography{references12}

%
%
%
%
%

%
%
\end{article}
\clearpage


%
%
%
%
%
%
%
%

\begin{figure}
\noindent\includegraphics[width=\textwidth]{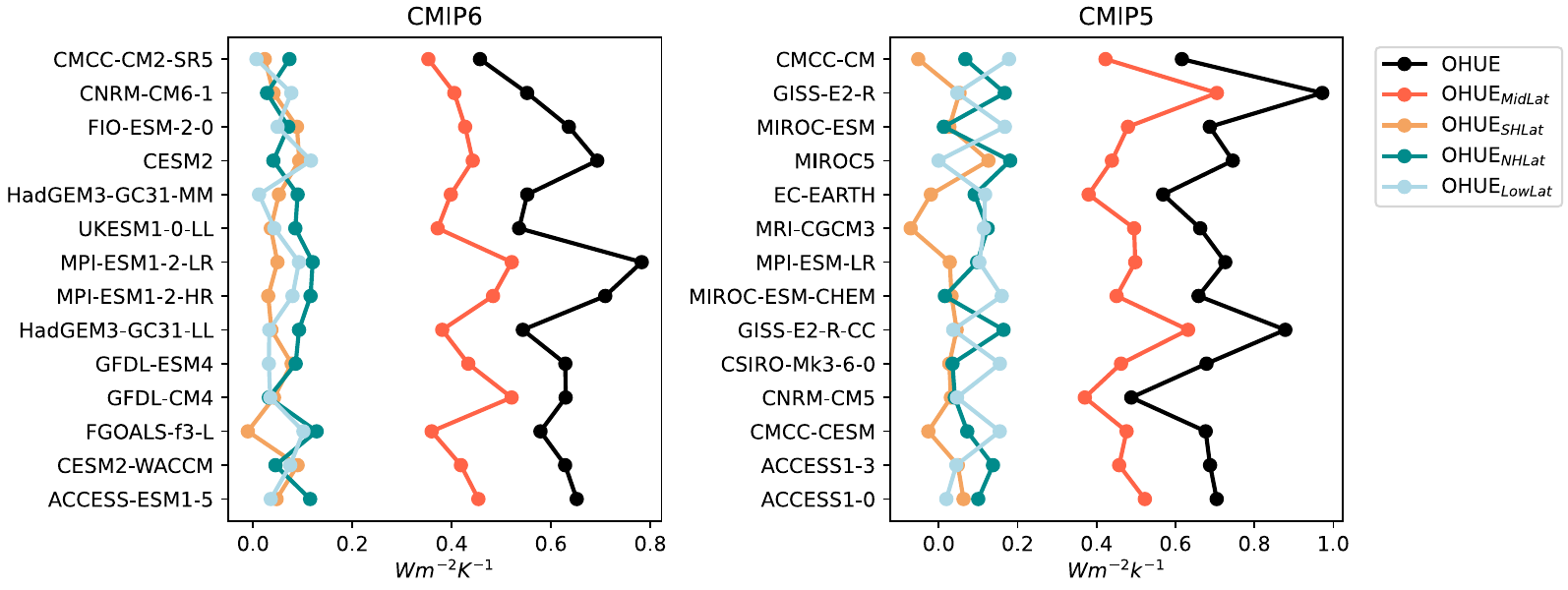}
\caption{Contribution of each regional OHUE component across models in CMIP6 (left)
and CMIP5 (right).
June}
\label{fs3}
\end{figure}

\begin{figure}
\noindent\includegraphics[width=\textwidth]{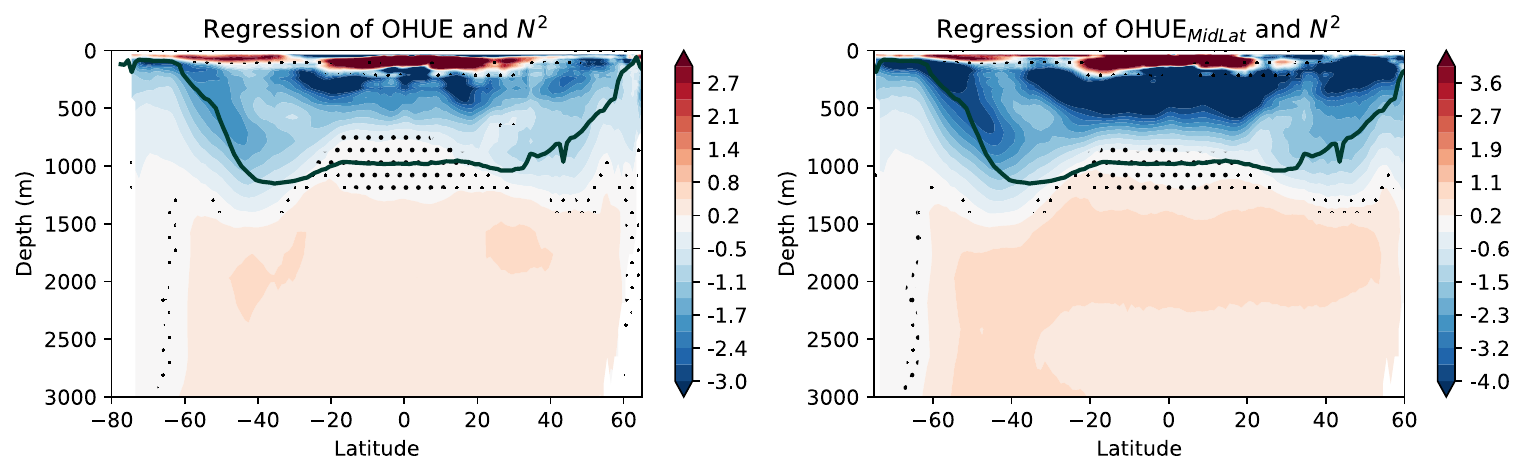}
\caption{The slope of the regression on $\overline{N^2}$ of OHUE (left) and OHUE$_{MidLat}$ (right). This
figure emphasizes strong differences in magnitude of vertical stratification in the upper and lower
ocean. Specifically, while there is a significant, positive correlation between $\overline{N^2}$ and both OHUE
and OHUE$_{MidLat}$ below ≈1500 m (see Figs. 3b and 4a in the main text), the regression show
that differences in deep ocean stratification between high and low OHUE models are extremely
small compared to the differences in the upper ocean.}
\label{fs4}
\end{figure}

\begin{figure}
\noindent\includegraphics[width=.8\textwidth]{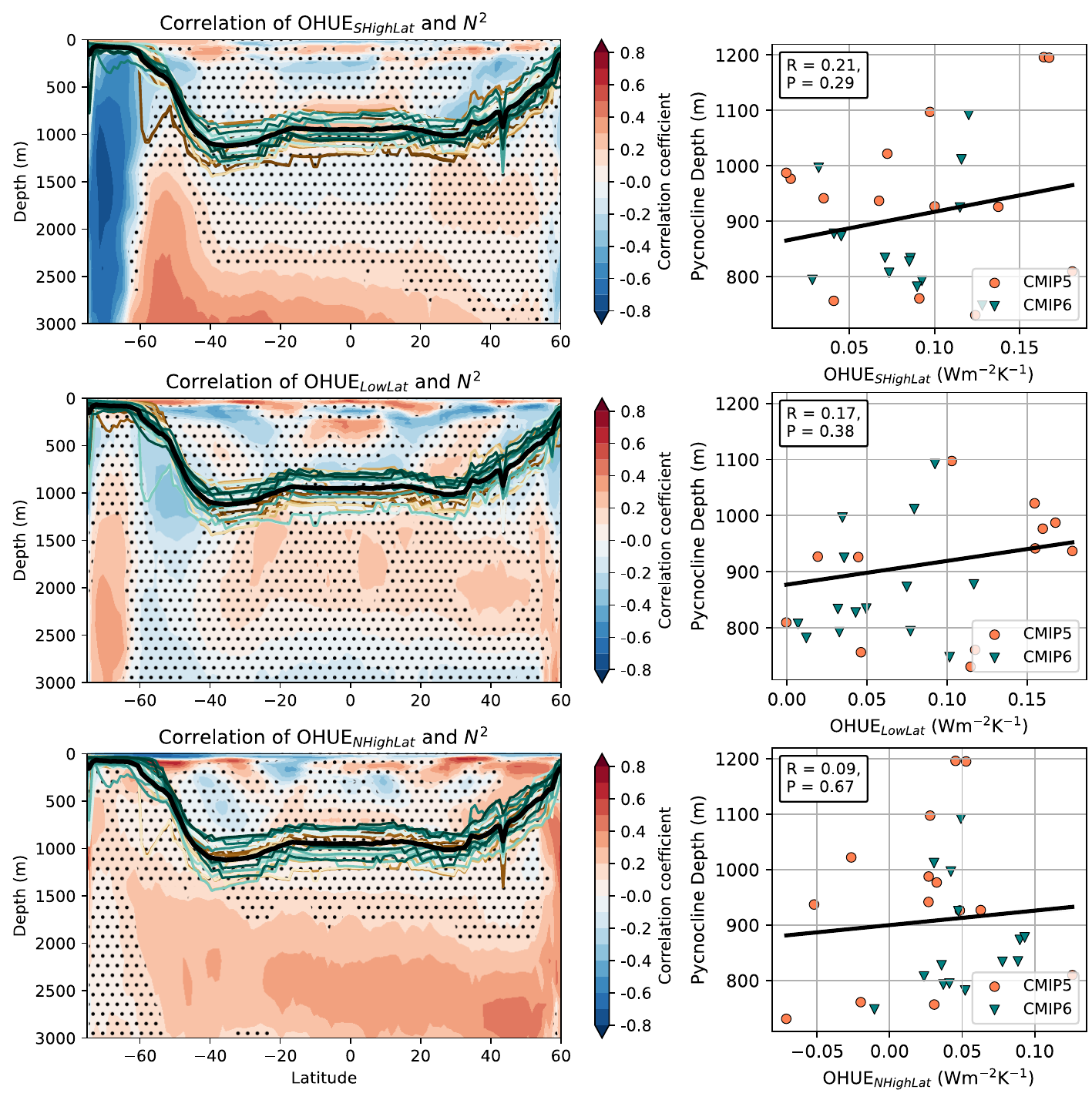}
\caption{Left column) Correlation of OHUE$_{SHighLat}$ (top), OHUE$_{LowLat}$ (middle), and
OHUE$_{NHighLat}$  (bottom) and $\overline{N^2}$ in CMIP5-6 Right column) Correlation of average subtropical
pycnocline depth and OHUE$_{SHighLat}$ (top), OHUE$_{LowLat}$ (middle) and OHUE$_{NHighLat}$ (bottom).
In all plots, the ensemble average pycnocline depth is overlaid in black, and overlaid in color
for each region, R, are the pycnocline depths in each model, colored from high (brown) to low
(green) OHUE$_{R}$}
\label{fs6}
\end{figure}

%
%
%
%
%